\documentclass[conference]{IEEEtran}
\IEEEoverridecommandlockouts

\usepackage{cite}
\usepackage{amsmath,amssymb,amsfonts}
\usepackage{algorithmic}
\usepackage{graphicx}
\usepackage{textcomp}
\usepackage{xcolor}
\usepackage{url}
\usepackage{booktabs} 
\usepackage{tabularx}
\usepackage{array}
\usepackage{enumitem} 
\usepackage[ruled,vlined]{algorithm2e}
\def\BibTeX{{\rm B\kern-.05em{\sc i\kern-.025em b}\kern-.08em
    T\kern-.1667em\lower.7ex\hbox{E}\kern-.125emX}}
\begin{document}

\title{ The Impact of Side Information on Voice Privacy
}

\author{
\IEEEauthorblockN{
Eulalie Thiombiano\textsuperscript{1,2},
Martha Larson\textsuperscript{1},
Vincent Colotte\textsuperscript{2},
Emmanuel Vincent\textsuperscript{2}
}
\IEEEauthorblockA{
\textsuperscript{1}Radboud University, Nijmegen, The Netherlands
\quad
\textsuperscript{2}Université de Lorraine, CNRS, Inria, LORIA, Nancy, France\\
eulalie.thiombiano@ru.nl
}
}

\maketitle

\begin{abstract}
Currently, voice-related privacy risks are studied under the assumption that the attacker has access only to spoken audio.
In contrast, real-world collections often pair the speech signal with side information; yet to our knowledge, no publicly available dataset provides sufficiently detailed and
directly linkable speech and side-information pairs to evaluate the combined
risk. We propose a distribution-constrained method to build a semi-synthetic
dataset and study singling out risk when geographic side information is
assumed to accompany anonymized speech. We instantiate it using the English
portion of Mozilla Common Voice and the 2021 UK Census, under ignorant and
semi-informed attackers, and introduce a conditional reference that isolates
the contribution of the audio beyond the candidate-set reduction induced by
the side information. In our controlled geographic case study, for
single-utterance speakers, the semi-informed attacker reaches an observed
isolation probability of 0.718 in the finest OA-plus-household setting,
compared with 0.380 in the signal-only setting; its excess over the
conditional reference increases from 0.012 to 0.340. The ignorant attacker
shows the same trend at a lower level. The effect is largest at fine
granularity and short recordings, precisely the setting in which signal-only
evaluation looks safest. Our results show that geographic side information can
increase the singling out risk of anonymized speech, and we recommend
including attacker settings with side information in future studies of voice
privacy.
\end{abstract}

\section{Introduction}
\label{sec:introduction}

Current voice anonymization benchmarks define anonymization mainly at signal level. The leading example is the VoicePrivacy Challenge~\cite{tomashenko2022voiceprivacy,10603395,tomashenko2024voiceprivacy,tomashenko2026third,tomashenko2025first}, which suppresses speaker identity through signal-level transformations based on voice conversion and speaker embeddings. Similarly, approaches with formal privacy guarantees remove speaker information from acoustic features~\cite{shamsabadi2022differentially}. Consequently, privacy evaluation is usually framed as signal-based re-identification risk, with metrics computed from speech, speech-derived representations, or similarity scores~\cite{maouche2020comparative,nautsch2020privacy,backstrom2025privacy}.

In practice, however, audio recordings may be released, analyzed, or deployed together with side information. Such information can be collected at recording time~\cite{ardila2020common}, inferred from acoustic features~\cite{yang2025demographic}, or obtained from external sources and linked to the recordings~\cite{hechmi2021voxceleb}. Under the GDPR, identifiability must be assessed by considering all means reasonably likely to be used to identify an individual~\cite{gdpr_2016,wp29_2014_anonymisation}. When speech data coexist with side information, these means may include combining the speech signal with the available information.
Signal-level evaluations that ignore this combination may underestimate privacy risk.

Evaluating this combined risk requires datasets that pair real speech with
side information beyond what the acoustic signal alone reveals. Existing public
corpora fall short of this need: they were designed for speech-processing tasks,
not privacy risk assessment, and their metadata are limited. Enriching them with
additional speaker attributes is problematic. For identifiable speakers, such
as those in VoxCeleb, attributes must be reconstructed from external sources,
and may be uncertain or unavailable~\cite{kacprzak2024heightceleb}; moreover,
centralizing dispersed personal information can itself violate privacy, even
when individual pieces are publicly accessible. For pseudonymized corpora such
as Common Voice, there is no public identity to link against. Directly
collecting such attributes would create the very exposure the evaluation aims
to study.

In this paper, we address this gap with a distribution-constrained approach that assigns semi-synthetic geographic attributes to speakers using population-level census statistics.
Geographic location is a suitable case study because it is commonly associated with user data, has a natural granularity, and can be modeled from public population statistics.
We instantiate the approach by augmenting the English portion of Mozilla Common Voice v11.0, the corpus used in~\cite{vauquier2025legally}, with UK 2021 Census-consistent geographic attributes.

Within this setting, we measure privacy risk through singling out, defined as the isolation of a unique record within a dataset. Following Cohen and Nissim~\cite{cohen2020towards}, we use Predicate Singling Out (PSO) to test whether predicates isolate individuals beyond chance.
Recent work applies PSO to anonymized speech and shows that it captures GDPR-oriented identifiability aspects not reflected by speaker-verification metrics such as the equal error rate (EER), but evaluates risk from the speech signal alone~\cite{vauquier2025legally}.
We build on this signal-only PSO setting and extend it to a new type of attacker who additionally observes a geographic attribute. The attacker's goal is not to recover a civil identity but to isolate a unique speaker by combining acoustic and geographic conditions.
We consider two attackers, ignorant and semi-informed. For each, we vary the granularity of the geographic side information, from coarse to fine\footnote{\url{https://github.com/Koguili/The-Impact-of-Side-Information-on-Voice-Privacy}}.

Our contributions are threefold: (i) we propose a distribution-constrained methodology for constructing  speech datasets containing side information without exposing speaker identity; (ii) we formalize a conditional reference isolation probability within the PSO framework to disentangle the effect of geographic pool restriction from that of acoustic matching; and (iii) using the resulting semi-synthetic dataset as a controlled case study, we show when geographic side information changes the conclusion of a signal-only PSO assessment [17], across geographic granularities and speech-session lengths.

Section~\ref{sec:importance} provides background on side information in speech applications and presents the PSO framework used in this work. Section~\ref{sec:methodo} describes the distribution-constrained methodology and its instantiation. Section~\ref{sec:experiment} describes the experimental setup, including the attacker assumptions and the evaluation framework. Section~\ref{sec:results} reports and discusses the singling out results. Section~\ref{sec:conclusion} concludes the paper and outlines future work.

\section{Background}
\label{sec:importance}
\subsection{Importance of side information}

Data are often task-specific, but audio recordings are rich and may contain information beyond the intended target content. This information may be embedded in the signal, stored as metadata, or derived from external sources linked to the recordings. We refer to all such elements as side information.

Side information is common in speech applications. Health studies combine speech~\cite{riad2024automated} or cough~\cite{kafentzis2023predicting} recordings with demographic and clinical data. Speech datasets use metadata to support interpretation~\cite{lee2022building,badawy2020metadata}. Similar uses appear in call centers with contextual annotations~\cite{feng2023end}, voice services using geolocation~\cite{chelba2015geo,setiawan2024implementation}, and voice biometrics studies of demographic disparities~\cite{chen2022exploring}.

Privacy work shows that such information can increase identifiability. Removing direct identifiers does not ensure anonymity: demographic attributes can uniquely identify many people~\cite{sweeney2000simple}.
High-dimensional, sparse datasets are vulnerable when combined with background knowledge, as shown for recommender systems, where behavioral fingerprints make anonymization fragile~\cite{narayanan2008robust,calandrino2011you}. Since speech data are high-dimensional and context-rich, similar risks may arise in voice settings, but they remain underexplored.

We therefore study geolocation as a representative form of side information. Speech can itself carry coarse location cues such as accent, language, and background sounds~\cite{chasmai2025audio,golle2009anonymity,backstrom2025privacyinspeechtechnology,williams2023new}. Rather than relying on such inference, we study the complementary and more direct case where a geographic attribute accompanies the recording, through collection or leakage. Following tabular privacy work, where generalizing quasi-identifiers is a standard way to reduce re-identification risk~\cite{fung2010privacy}, we vary geographic granularity to assess its effect on speech privacy risk.

\subsection{Predicate Singling Out}
\label{sec:pso}
We use the Predicate Singling Out (PSO) framework of Cohen and Nissim~\cite{cohen2020towards}. Let $X$ denote the data universe and let $D$ be a population distribution over $X$. A dataset is an $n$-tuple
\begin{equation}
x = (x_1,\ldots,x_n) \sim D^n,
\end{equation}
where each component $x_i \in X$ is a row of the dataset. Each row represents the full data associated with one individual, who appears at most once. A predicate is a binary function $p : X \rightarrow {0,1}$. In the PSO framework, a predicate isolates a row, and therefore one individual, in $x$ when
\begin{equation}
\sum_{i=1}^{n} p(x_i) = 1.
\label{eq:isolation_condition}
\end{equation}

The population weight of the predicate is
\begin{equation}
w \triangleq \mathrm{wt}_D(p) = \Pr_{z \sim D}[p(z)=1],
\end{equation}
i.e., the probability that a randomly drawn individual satisfies $p$. Under the i.i.d.\ sampling model $x \sim D^n$, the variables $p(x_1),\ldots,p(x_n)$ are independent Bernoulli random variables with parameter $w$, so that
\begin{equation}
\sum_{i=1}^{n} p(x_i) \sim \mathrm{Binomial}(n,w).
\end{equation}

The isolation condition therefore corresponds to the binomial event of exactly one success, whose probability is
\begin{equation}
\Pr_{x\sim D^n}[\mathrm{iso}(p,x)] = \binom{n}{1} w (1-w)^{n-1} = n w (1-w)^{n-1}.
\label{eq:baseline}
\end{equation}
We refer to Eq.~\eqref{eq:baseline} as the \emph{baseline}: the isolation probability expected purely from random sampling for a predicate of weight $w$ in a dataset of size $n$. It is the level of isolation that can occur by chance, without exploiting dataset-specific information.

The baseline is maximized when the predicate weight is $w=1/n$, which corresponds to an expected single match, since $nw=1$. Substituting $w=1/n$ into Eq.~\eqref{eq:baseline} gives
\begin{equation}
n w (1-w)^{n-1} =
\left(1 - \tfrac{1}{n}\right)^{n-1}
\longrightarrow e^{-1} \approx 0.37
\quad \text{as } n \to \infty .
\end{equation}
This limit gives the order of magnitude of isolation attainable by chance alone. Prior signal-only evaluation adopts this asymptotic value as a fixed reference~\cite{vauquier2025legally}: isolation above 0.37 is read as evidence of an acoustic attack. We return to this choice in~\ref{sec:conditionalBaseline}, where the availability of side information makes a fixed reference inappropriate and motivates a context-dependent one.

\section{Methodology}
\label{sec:methodo}
\subsection{From data integration to speech datasets containing side information}
\label{sec:motivation}

Our goal is to test whether signal-only evaluations capture privacy risk when an attacker has side information about speakers. This requires pairing real speech with such information, but the relevant variables are stored in separate sources and are never jointly observed for the same speakers.

This is a data integration problem~\cite{gao2017data,rubin1986statistical,lewaa2021data}: record linkage connects records from the same individuals, while statistical matching combines sources through shared variables when direct linkage is impossible. Since variables never jointly observed at individual level cannot be recovered from the data alone, any constructed joint dataset relies on assumptions or explicit distributional constraints~\cite{d2024state}. Namazi-Rad et al.~\cite{namazi2017unconstrained} provide a useful precedent: they build a geo-specific synthetic population from census and survey data by matching individuals and households to census margins and imputing survey-derived attributes. We adopt the principle: when individual-level data are unavailable, population-level constraints can yield micro-level data validated by distributional fidelity.

Our setting differs because speech recordings and embeddings remain real, assigned geographic attributes are not true locations, and assignment is performed over population categories rather than identifiable individuals.
The resulting dataset is therefore semi-synthetic: real speech paired with geographic side information under population-level constraints. Since speech and geographic attributes are not jointly observed for the same speakers, an assignment step is required to construct such a dataset.
\subsection{Distribution-constrained geographic construction}
\label{subsec:approach}

Building on the incomplete-data perspective above, we construct a speech dataset containing side information by assigning population-consistent geographic attributes to real speakers. Let $\mathcal{D}_{\mathrm{sp}}$ be a speech corpus with unique speakers $\mathcal{S}={s_1,\ldots,s_M}$, where each speaker $s$ has an audio signal $A(s)$ and possibly metadata. Let $\mathcal{D}_{\mathrm{ext}}$ be an external population-level source. We instantiate the construction with the English portion of Mozilla Common Voice v11.0 ($M=22{,}024$ speakers), the 2021 UK Census, and Rutland (LAD E06000017) as a controlled geographic case study.

\paragraph{Side information.}
We model an attacker who observes, in addition to the released audio, a side information $Q$ with category set $\mathcal{Q}$. An attribute qualifies as attacker-accessible side information when it reduces uncertainty about speaker identity beyond what the speech signal alone reveals. This does not require $Q$ to be independent of speech: an attribute partly inferable from audio may still add information when observed more reliably, at finer granularity, or in a form not directly recoverable from the signal. In this work, $Q$ is the assigned geographic attribute, evaluated from coarse to fine granularity: Local Authority District (LAD) $\supset$ Middle Layer Super Output Area (MSOA) $\supset$ Lower Layer Super Output Area (LSOA) $\supset$ Output Area (OA). We also generate synthetic household identifiers within OAs to group speakers into one-person, two-person, and three-or-more-person households; only the resulting \emph{household size category} (HH1, HH2, or HH3+), not the individual household identifier, is disclosed as side information, further refining the OA-level attribute. At the LAD level, all speakers share the same value, so the setting coincides with signal-only evaluation. At finer levels, $Q$ partitions speakers into smaller compatible groups, which determine the conditional reference in~\ref{sec:conditionalBaseline}.

\paragraph{Constraint variables.} Variables $C$, observed in $\mathcal{D}_{\mathrm{sp}}$ and available in aggregate form in $\mathcal{D}_{\mathrm{ext}}$, are used only to calibrate the assignment. They are not disclosed to the attacker. This separates construction variables from attacker-accessible side information. In our instantiation, $C=(\mathrm{gender},\mathrm{age\ group})$: speakers assigned to an OA should match its census age-by-gender composition in distribution, not at the individual level.

\paragraph{Assignment.}
Rather than linking speakers to real individuals, we define an assignment function
\begin{equation}
\phi:\mathcal{S}\to\mathcal{Q},
\label{eq:assignment_function}
\end{equation}
chosen so that the assigned attributes match the external distribution, both marginally and conditionally on $C$:
\begin{equation}
\hat{P}_{\phi(\mathcal{S})}(Q=q)
\approx
P_{\mathcal{D}{\mathrm{ext}}}(Q=q),
\label{eq:assignment_marginal}
\end{equation}
\begin{equation}
\hat{P}_{\phi(\mathcal{S})}(Q=q \mid C=c)
\approx
P_{\mathcal{D}_{\mathrm{ext}}}(Q=q \mid C=c).
\label{eq:conditional_assignment}
\end{equation}
These are design targets, not assumptions about the speakers' true attributes. The signal $A$ is not used in the assignment, so the construction introduces no acoustic--geographic dependence beyond what is captured by $C$.

The procedure uses the speaker set $\mathcal{S}$, OA-level adult population counts, the OA--LSOA--MSOA--LAD lookup table, OA-level age-by-gender distributions from Census Table TS009, and OA-level household composition from Census Table TS003\cite{nomis}. Full details are available in the project repository.

Speakers are first allocated to OAs in proportion to population:
\begin{equation}
\alpha_i =
\frac{\mathrm{pop}_{i}}
{\sum_j \mathrm{pop}_{j}},
\qquad
\tilde{n}_i = \alpha_i M .
\label{eq:oa_proportions}
\end{equation}
Since $\tilde{n}_i$ is generally fractional, largest-remainder apportionment~\cite{brill2018multiwinner} gives integer OA counts $n_i$ satisfying
\begin{equation}
\sum_i n_i = M .
\label{eq:oa_sum}
\end{equation}
Thus, OA counts follow the census-derived population distribution by construction. The OA marginal distribution is therefore not an empirical assumption but a construction constraint: after largest-remainder apportionment, each assigned OA count differs from its census-proportional target by at most one speaker. Any remaining mismatch concerns only the secondary constraints, namely the age--gender slots and the synthetic household grouping.
Within each OA, speakers are assigned to age-by-gender slots from TS009, rescaled to the OA speaker count. Household structure is then added from TS003 by grouping speakers into synthetic one-person, two-person, and three-or-more-person households. Members of the same synthetic household share both an OA and a household identifier. LSOA, MSOA, and LAD labels are then derived from the OA lookup table. Each speaker keeps their real recordings; only the geographic and household attributes are synthetic.

\paragraph{Assumptions and generality.}
The construction assigns geography conditionally on demographic constraints, but without direct dependence on acoustic content. The $22{,}024$ Common Voice speakers are matched to Rutland-derived population structure, without claiming that they are representative of Rutland residents. Assigned attributes reproduce external distributions but are not true residences or household memberships. Household identifiers model co-location within an OA, without constraining household-internal demographics.

\subsection{A conditional reference probability for evaluation with side information}
\label{sec:conditionalBaseline}

The construction in~\ref{subsec:approach} defines the dataset. We now adapt the
PSO reference of~\ref{sec:pso} to the disclosed side information: for each
target, isolation must be assessed within the set of candidates sharing the
disclosed value.

\paragraph{Compatible block.}
For each evaluation draw, let $\mathcal{S}_N \subset \mathcal{S}$ be the
candidate set of size $N$. The compatible block of target speaker $t$ is defined as:
\begin{equation}
G_N(t)=\{\,s\in\mathcal{S}_N : Q(s)=Q(t)\,\}, \qquad n'_t=|G_N(t)|.
\label{eq:compatible-block}
\end{equation}
We write $G(t)$ for $G_N(t)$ when the draw is clear; with multiple disclosed attributes, the block is defined by their joint values.
The attacker's predicate is
\begin{equation}
p_t(s) = \mathbf{1}\{f_t(s)>\theta\}\cdot\mathbf{1}\{s\in G(t)\},
\label{eq:side-information-predicate}
\end{equation}
where $f_t(s)$ is the acoustic similarity between target $t$ and candidate $s$,
and $\theta$ is calibrated within $G(t)$ so that one candidate is expected to
match (\ref{sec:evaluation}).

\paragraph{Conditional reference.}
Vauquier et al.~\cite{vauquier2025legally} use a fixed PSO reference close to
$e^{-1}\approx0.37$. With side information, this reference must depend on the
compatible block size. Since the threshold is calibrated to one expected match,
$w_t\approx 1/n'_t$, giving
\begin{equation}
P_{\mathrm{ref}}(n'_t)
=
\left(1-\frac{1}{n'_t}\right)^{n'_t-1}.
\label{eq:conditional-reference-probability}
\end{equation}
We define $\Delta_{\mathrm{iso}}=P_{\mathrm{obs}}-P_{\mathrm{ref}}(n'_t)$:
positive values indicate acoustic contribution within the compatible pool,
while values near zero indicate isolation explained by pool restriction.

Eq.~\eqref{eq:conditional-reference-probability} assumes independent acoustic
matches within a block. Since geography is assigned without using the signal
(\ref{subsec:approach}), the construction introduces no direct
acoustic--geographic dependence beyond the constraints $C$. We discuss possible
departures from this assumption in~\ref{subsec:discussion}.

\paragraph{Regimes and generality.}
The reference depends only on $n'_t$: it is close to $e^{-1}$ for large blocks,
higher for small blocks, and equals $1$ when $n'_t=1$, where side information
alone isolates the speaker. This prevents isolation due only to candidate-set
reduction from being counted as an acoustic attack.

\section{Experimental setup}
\label{sec:experiment}

\subsection{Specification of the attacker}
\label{sec:attacker}
We specify the attacker by its objective, opportunity, and resources, following~\cite{rahman2024scenario}. All released recordings are anonymized with the VoicePrivacy 2024 B1 system\cite{voiceprivacy2024github}, based on x-vector modification and neural source-filter synthesis, in the utterance-level setting.
Both attackers aim to construct a predicate that isolates exactly one individual. Both access anonymized recordings and disclosed geographic side information at MSOA, LSOA, and OA granularities, further refined by household size categories. LAD is not an additional resource because all speakers share the same LAD value, reducing it to the signal-only setting.
The attackers differ in their resources. The \emph{ignorant attacker}
uses the SIDEKIT speaker-embedding extractor, trained on non-anonymized speech and not adapted to B1. The \emph{semi-informed attacker} knows the anonymization mechanism: we anonymize LibriSpeech train-clean-360 with B1, train an ECAPA-TDNN extractor on the resulting speech, and use it to compute embeddings from the anonymized Common Voice recordings for the singling out evaluation in~\ref{sec:evaluation}.

\subsection{Evaluation framework}
\label{sec:evaluation}

\paragraph{Full-disclosure protocol.}
To evaluate singling out risk, we follow the protocol of~\cite{vauquier2025legally},
which operationalizes the PSO framework of~\cite{cohen2020towards}
(\ref{sec:pso}): predicates are built from acoustic similarity between speaker
embeddings, isolation is evaluated through Eq.~\eqref{eq:isolation_condition},
and we extend the predicate with geographic side information as defined
in~\ref{sec:conditionalBaseline}. The similarity is computed with the attacker's
embedding extractor (\ref{sec:attacker}), so the same protocol applies to both
attackers with their respective extractors.

Following~\cite{vauquier2025legally}, Common Voice utterances are split into
two utterance-disjoint subsets, $A$ and $B$, containing $22{,}024$ and $4{,}949$
speakers, respectively, with speakers in $B$ being a subset of those in $A$.
For each conversation length $L$, candidate enrollment speakers are taken from
$B$ and must have at least $10L$ utterances in $A \cup B$; we randomly sample
$495$ of them for enrollment. Evaluation datasets $\mathcal{S}_N$ of size $N$
are sampled from $A$. For each evaluation speaker, $10L$ utterances are split
into $9L$ calibration and $L$ test utterances.

For each target $t$ from the enrollment speakers set, we restrict the current evaluation set $\mathcal{S}_N$ to the compatible block $G(t)$, leaving $n'_t=|G(t)|$ compatible candidates. The
acoustic predicate $p_\theta(s)=\mathbf{1}\{f_t(s)>\theta\}$ compares target
$t$'s enrollment representation to candidate $s$ from test set, and $\theta$ is calibrated
within $G(t)$ so that one compatible candidate is expected to match. Combined
with the block restriction, this gives the side-information predicate of
Eq.~\eqref{eq:side-information-predicate}. The singling out score per target is the
percentage of enrollment comparisons for which this predicate yields exactly
one match among the test speakers.

We vary the evaluation size $N$ and the conversation length $L$; since each
evaluation speaker must provide $10L$ utterances, the maximum $N$ decreases as
$L$ grows. For each $L$, we repeat the evaluation over five random draws of
evaluation speakers and ten calibration/test splits per draw, yielding
$495\times 5\times 10 = 24{,}750$ enrollment evaluations per configuration.
Each configuration is compared to its conditional reference
(Eq.~\eqref{eq:conditional-reference-probability}), evaluated at each run's
block size $n'_t$ rather than at $N$, since the reference is nonlinear in
$n'_t$.

\paragraph{Partial side information.}
We further evaluate settings where the geographic attribute is disclosed only
for a fraction $\pi \in [0,1]$ of the candidates in $\mathcal{S}_N$
(\emph{known\_frac}), drawn uniformly at random and independently of $C$.
For a target $t$ with disclosed location, the compatible block is
\[
\begin{aligned}
G_\pi(t)=\ &\{s\in\mathcal{S}_N: Q(s)=Q(t)\} \\
\cup\ &\{s\in\mathcal{S}_N: Q(s)\ \text{is undisclosed}\}.
\end{aligned}
\]
A disclosed candidate is kept only if it matches $Q(t)$; an undisclosed one is
kept because it cannot be excluded. The threshold $\theta$ is calibrated within
$G_{\pi}(t)$ as above. We report total isolation within $G_{\pi}(t)$, decomposed
by whether the isolated candidate's attribute was disclosed
(\emph{known}) or not (\emph{unknown}).

\section{Results}
\label{sec:results}

\begin{figure*}[t]
\centering
\includegraphics[width=\textwidth]{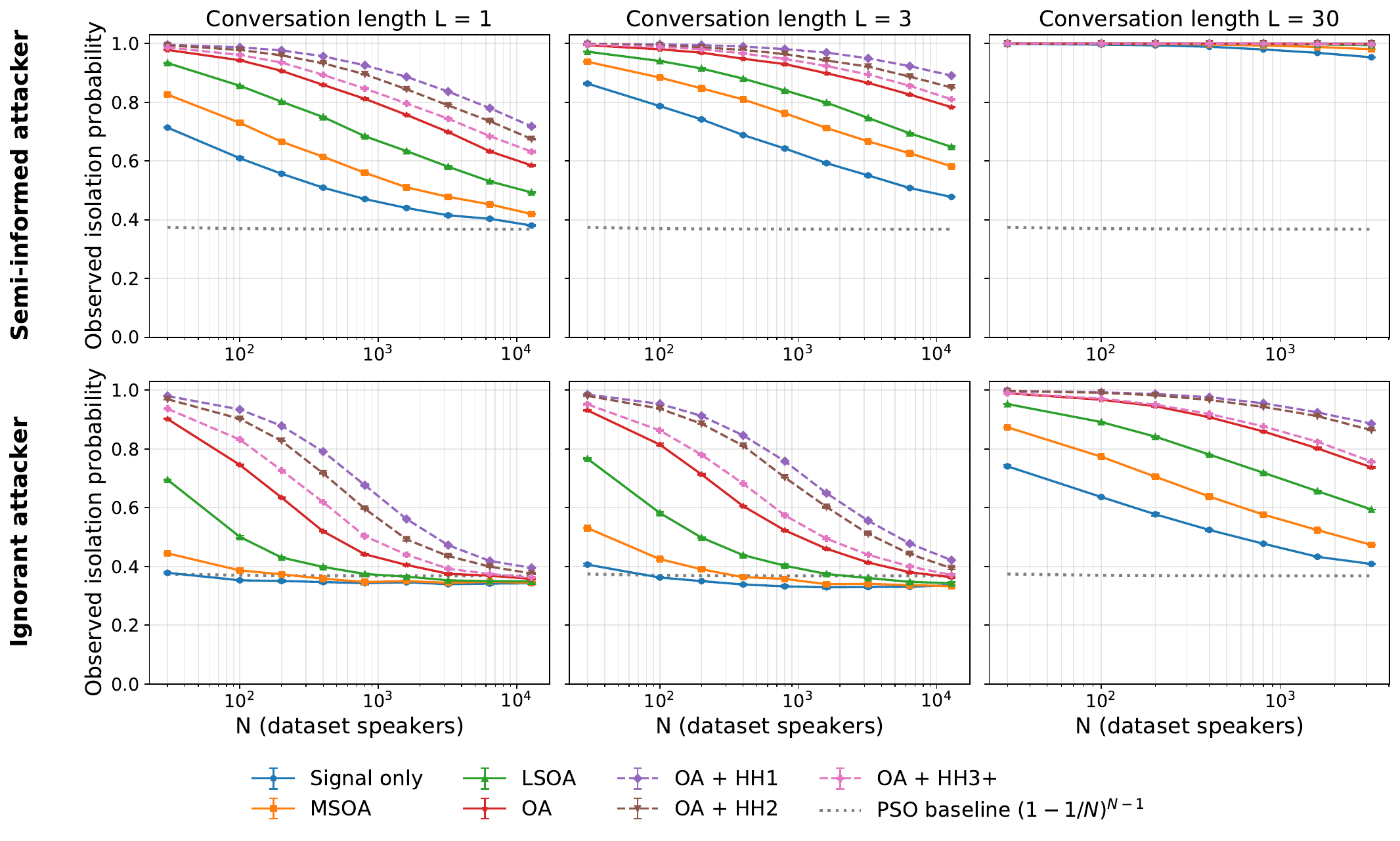}
\caption{Observed isolation probability for the semi-informed (top) and ignorant (bottom) attackers. Error bars show the SEM across runs and are smaller than the markers.}
\label{fig:results_combined}
\end{figure*}

\subsection{Observed isolation with side information}
\label{sec:obs}

\paragraph{Granularity and assignment robustness.}
Figure~\ref{fig:results_combined} shows that observed isolation increases as side information becomes finer: signal only $<$ MSOA $<$ LSOA $<$ OA. Synthetic households further refine OA-level information, with OA+HH1 $>$ OA+HH2 $>$ OA+HH3+, because smaller households induce smaller compatible blocks. This pattern holds for both attackers. The semi-informed attacker reaches higher values because its extractor is adapted to B1, whereas the ignorant attacker remains lower but follows the same trend. As $L$ increases, the speech signal becomes more discriminative and the marginal effect of side information decreases.

These trends are not driven by the assignment constraints. OA only, OA+age, OA+gender, and OA+gender+age give almost identical isolation probabilities, with the largest difference below $0.003$ over all configurations, e.g.\ $0.699$ vs.\ $0.701$ at $L=1$, OA, and $N=3200$. We therefore use the joint OA--gender--age assignment, which best preserves the population structure; gender and age are used only for construction and are never disclosed to the attacker. This confirms that the reported increase in isolation is driven by the disclosed geographic partition.

\subsection{Contribution beyond candidate-set reduction}
\label{sec:contribution}

Observed isolation alone does not prove a successful acoustic attack, since side information can isolate by shrinking the candidate pool. We therefore compare $P_{\mathrm{obs}}$ with the conditional reference $P_{\mathrm{ref}}$ of~\ref{sec:conditionalBaseline} and define
\begin{equation}
\Delta_{\mathrm{iso}}
=
P_{\mathrm{obs}} - P_{\mathrm{ref}}(n'_t).
\label{eq:delta_iso}
\end{equation}
The difference is computed per run using the observed compatible block sizes, since $P_{\mathrm{ref}}$ is nonlinear in $n'_t$; Table~\ref{tab:summary_highlights} reports run-averaged values.

For $L=1$ and $N=12{,}800$, OA side information reduces the average compatible block to $\bar{n}'_t\approx123$, with $P_{\mathrm{ref}}=0.370$, while $P_{\mathrm{obs}}=0.585$. Thus, $\Delta_{\mathrm{iso}}=0.215$: geography narrows the search space, but the voice still contributes within it. The contribution is largest for OA+HH1, with $\Delta_{\mathrm{iso}}=0.340$. For the ignorant attacker, signal-only gives a negative contribution, $\Delta_{\mathrm{iso}}=-0.025$, but OA+HH1 makes it positive, $\Delta_{\mathrm{iso}}=0.017$. A setting ineffective under signal-only evaluation can therefore become effective once side information is available.

\begin{table}[t]
\centering
\caption{Selected results for $L=1$, $N=12800$. $\bar{n}'_t$: mean compatible block size; $P_{\mathrm{ref}}$: conditional reference; $P_{\mathrm{obs}}$: observed isolation; $\Delta_{\mathrm{iso}}$: contribution beyond the reference.}
\label{tab:summary_highlights}
\resizebox{\linewidth}{!}{%
\begin{tabular}{llrrrr}
\toprule
Attacker / setting & Side information & $\bar{n}'_t$ & $P_{\mathrm{ref}}$ & $P_{\mathrm{obs}}$ & $\Delta_{\mathrm{iso}}$ \\
\midrule
Semi-informed & Signal only & 12800.0 & 0.368 & 0.380 & 0.012 \\
Semi-informed & MSOA        & 2582.7  & 0.368 & 0.419 & 0.052 \\
Semi-informed & LSOA        & 540.3   & 0.368 & 0.493 & 0.125 \\
Semi-informed & OA          & 123.3   & 0.370 & 0.585 & 0.215 \\
Semi-informed & OA+HH1      & 24.0    & 0.378 & 0.718 & 0.340 \\
Semi-informed & OA+HH2      & 36.8    & 0.374 & 0.675 & 0.301 \\
Semi-informed & OA+HH3+     & 77.5    & 0.372 & 0.632 & 0.260 \\
\midrule
Ignorant & Signal only & 12800.0 & 0.368 & 0.343 & -0.025 \\
Ignorant & OA+HH1      & 24.0    & 0.378 & 0.395 & 0.017 \\
\bottomrule
\end{tabular}%
}
\end{table}

\begin{figure}[t]
\centering
\includegraphics[width=\linewidth]{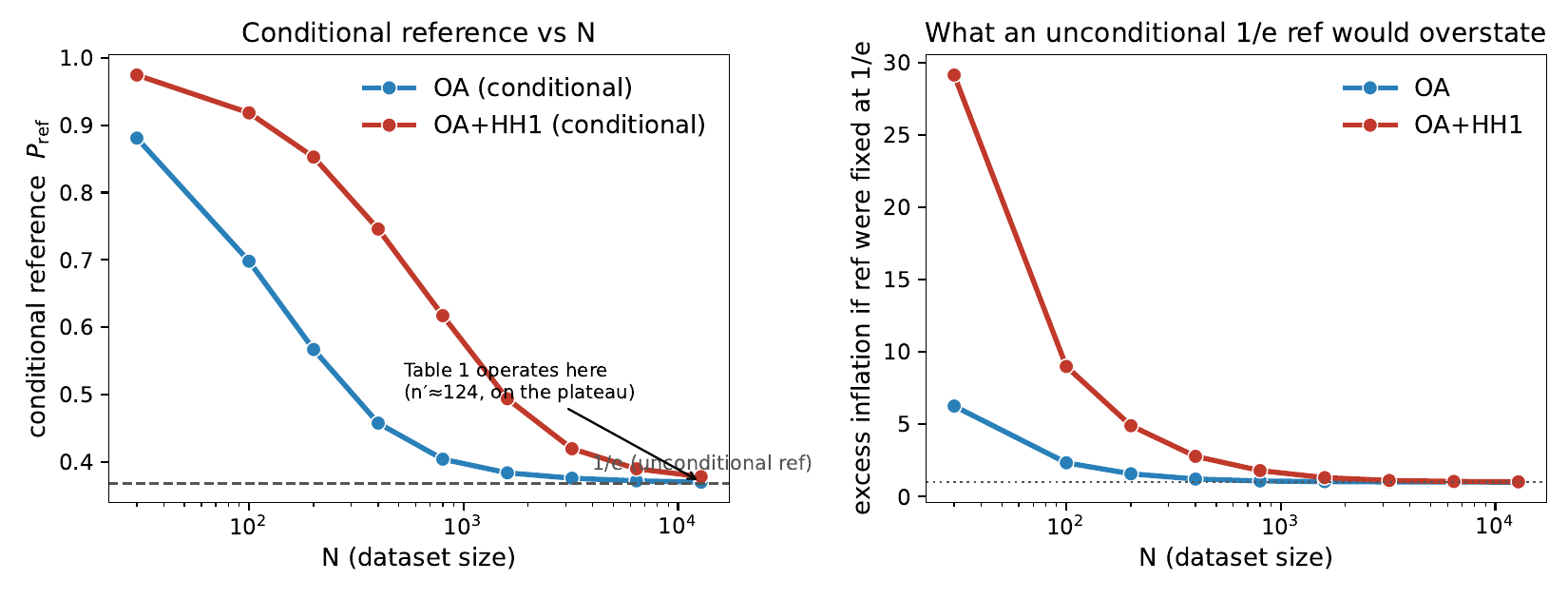}
\caption{Conditional reference $P_{\mathrm{ref}}$ for $L=1$. Left: $P_{\mathrm{ref}}$ decreases towards $e^{-1}$ as $N$ grows. Right: factor by which an unconditional reference fixed at $e^{-1}$ would overstate $\Delta_{\mathrm{iso}}$.}
\label{fig:conditional_reference}
\end{figure}

Figure~\ref{fig:conditional_reference} shows why a fixed reference is misleading. As $N$ grows, compatible blocks grow and $P_{\mathrm{ref}}$ approaches $e^{-1}\approx0.37$; for very small blocks it is much higher, reaching $0.88$ at $N=30$ and OA level, where $\bar{n}'_t\approx1.3$. A fixed $e^{-1}$ reference would overstate $\Delta_{\mathrm{iso}}$, by up to sixfold at OA level and about thirtyfold for single-person households.

\subsection{Effect of attacker knowledge, partial knowledge, length, and size}
\label{sec:knowledge}

\paragraph{Attacker and partial knowledge.}
The semi-informed attacker consistently outperforms the ignorant attacker because its extractor is trained on B1-anonymized speech, but side information helps both. For the ignorant attacker at $L=1$ and $N=12{,}800$, signal-only isolation is below the PSO reference ($0.343$ vs.\ $0.368$), while OA+HH1 raises it to $0.395$; each value averages $24{,}750$ enrollment evaluations, with SEM on $\Delta_{\mathrm{iso}}$ below $0.004$.

Even partial side information increases risk. When the attacker knows
locations for only $10\%$ of speakers ($L=1$), total observed isolation is
already above the signal-only level. The gain is concentrated on speakers
with a known location, showing that side information need not cover the whole
dataset to increase singling out risk.

\paragraph{Conversation length and dataset size.}
Conversation length controls signal strength. At $N=3200$, the semi-informed
signal-only isolation rises from $0.42$ at $L=1$ to $0.55$ at $L=3$ and $0.95$
at $L=30$, leaving little room for side information at long lengths. For the
ignorant attacker, however, fine-grained side information remains useful.
Increasing $N$ lowers isolation at short lengths by enlarging the candidate
pool, but the effect weakens as $L$ grows. Signal-only evaluation
underestimates risk most when acoustic evidence is weak.
\subsection{Discussion}
\label{subsec:discussion}
Our construction assigns location without using the speech signal and therefore
does not exploit voice--location correlations such as accent or background.
The measured risk should thus be interpreted as a controlled estimate under
correct and available side information. If such correlations remain after anonymization, they may
provide additional evidence to an attacker; if they are removed, the setting
approaches the independence assumed here.

The conditional reference depends on side information only through the
compatible block size induced by the disclosed attribute
(\ref{sec:conditionalBaseline}). It can therefore be applied beyond
geography to any quasi-identifier that narrows the candidate pool, such as
occupation, device, or recording context. Rutland is a controlled
instantiation rather than a representative setting: in another area, block
sizes may change, but the same reference would still separate candidate-set
reduction from within-block acoustic contribution.

It is important to note that age and gender only guide the assignment (cf. \ref{subsec:approach});
any use the attacker of these attributes is due to residual cues within the audio itself. 
Further, in this work the household category (i.e., HH1, HH2, or HH3+ label indicating the size of the household the speaker belongs to) is used for convenience as a category below the OA level. 
In future work on defense, households as a location constraint may become relevant.

\section{Conclusion}
\label{sec:conclusion}
Using the PSO framework, we built a semi-synthetic dataset pairing real
anonymized speech with census-consistent geographic attributes, and introduced
a conditional reference that separates candidate-set reduction from the
within-block voice contribution. Geographic side information increases
singling out risk, especially at fine granularities and short recordings,
where signal-only evaluation appears safest.

These results show that voice-privacy risk is not determined by the speech
signal alone. We therefore recommend reporting both signal-only settings and settings in which the attacker has access to side information when evaluating released audio. Future
work will relax the signal--geography independence assumption and also extend the
framework to other forms of side information.
\section{Acknowledgment}
This work was funded by the European Union’s Horizon Europe
research and innovation programme grant No 101168193.

Experiments presented in this paper were carried out using the Grid'5000 testbed, supported by a scientific interest group hosted by Inria and including CNRS, RENATER and several Universities as well as other organizations (see https://www.grid5000.fr).

\section{Generative AI Use Disclosure}

Generative AI tools were used for language editing and polishing, including grammar and phrasing. All scientific content, experimental design, analyses, results, and conclusions were developed, verified, and approved by the authors. The authors take full responsibility for the content of this paper, and no generative AI tool is listed as a co-author.

\bibliographystyle{IEEEtran}
\bibliography{references}
\end{document}